\documentclass[11pt]{article}
\usepackage[top=23.4mm, bottom=23.4mm, left=23.4mm, right=23.4mm]{geometry}
\usepackage{graphicx,dcolumn,bm}
\usepackage{amssymb,amstext,amsmath}

\usepackage{graphicx}
\usepackage{dcolumn}
\usepackage{bm}
\usepackage{amssymb}
\usepackage{multirow}
\usepackage{bigstrut}
\usepackage{makecell,rotating}
\usepackage{mathrsfs}
\usepackage{booktabs}
\usepackage{threeparttable}
\usepackage{multirow}
\usepackage{subfigure}
\usepackage{colortbl}
\usepackage{epsfig}
\usepackage{threeparttable}
\usepackage{chngpage}
\usepackage{float}
\usepackage{amstext}
\usepackage{amsmath}
\usepackage{amsthm}
\usepackage{pdflscape}
\usepackage{authblk}
\usepackage{hyperref}
\usepackage{rotating}
\usepackage[graphicx]{realboxes}
\usepackage{adjustbox}
\usepackage{setspace}
\usepackage{caption}
\usepackage{xcolor}
\usepackage{xr}
\usepackage{caption}
\usepackage{lineno}
\allowdisplaybreaks
\usepackage{enumitem}
\setitemize[1]{itemsep=1pt,partopsep=0pt,parsep=\parskip,topsep=4pt}
\title{Evolutionary hypergame dynamics: Introspection reasoning and social learning 
}
\author{Feipeng Zhang$^{1,2}$, Te Wu$^{1}$, Guofeng Zhang$^{2*}$ and Long Wang$^{3*}$}

\date{
    $^1$ Center for Complex Systems, Xidian University, Xi’an 710071, China\\
    $^2$ Department of Applied Mathematics, The Hong Kong Polytechnic University, Kowloon 999077, Hong Kong, China\\
    $^3$  Center for Systems and Control, College of Engineering, Peking University, Beijing 100871, China\\%
    $^*$ \footnotesize Corresponding authors. E-mail: guofeng.zhang@polyu.edu.hk,  longwang@pku.edu.cn\\[2ex]
}

\begin{document}
\maketitle
\begin{abstract}
In the realm of evolutionary game theory, standard frameworks typically presuppose that every player possesses comprehensive knowledge and unrestricted access to the entire  strategy space. However, real-world human society inherently harbors diverse levels of knowledge, experience, and background among individuals. Hypergames incorporate this heterogeneity by permitting individuals to differ in their access to the full strategy set, reflecting cognitive or informational constraints and giving rise to asymmetric strategic interactions. Yet, their evolutionary consequences remain underexplored. Our inquiry employs prototype models featuring three available strategies, focusing on social dilemmas involving cooperation, defection, and loner. These strategies manifest cyclic dominance, akin to the well-studied rock-paper-scissors dynamics, a foundational model in game theory.  Our study spans both well-mixed and spatial lattice populations, delving into the intricacies of learning and evolution of the strategy set within the evolutionary hypergame dynamics. In stark contrast to traditional evolutionary game dynamics, our findings unveil nuanced and intricate phases, encompassing scenarios of loner dominance, coexistence of multiple strategy sets, combinations of cooperation and loner dominance, and more. Remarkably, we discern that heightened rationality significantly promotes cooperative behaviors.
\end{abstract}

\section{Introduction}
Individuals adapt their behaviors to changing environments\cite{PNASHumansadapt,Naivelearningexperiments,PNAS_Traulsen2010,David_2013,YeWang_NSR}, and such adaptive processes have been widely examined across fields ranging from microbial dynamics\cite{Evolutionary_microbial} to human social systems\cite{Axelrod1981_science,Evolutionarycultures,ChibaPNAS2024Sociallearning}. Yet, quantifying how individuals learn, decide, and interact strategically remains a formidable challenge. Evolutionary game theory offers a well-founded framework for investigating behavioral adaptation\cite{Shenganzhi2024,guocheng2024,PNASTraulsen2010,JTBTarnita2009,Intreface_LaPorte2023}, built on the premise that individuals adjust gradually over time through mechanisms such as cultural transmission or genetic evolution\cite{chenfang_interface,leiNC2021}. 
When self-interested individuals interact, their pursuit of personal gain frequently produces outcomes detrimental to all parties, making the emergence and persistence of cooperation a central problem in the fields\cite{Nowak1993:WSLS,nature_any_structure2017,SuPNAS2019,NCenvironmentfeedbacks}.

To characterize how individuals acquire new behaviors, researchers have explored various cognitive mechanisms, including peer imitation\cite{FudenbergImitationprocesses}, aspiration-based learning\cite{ZhouAspiration,ChenPRE2008}, reinforcement learning\cite{McAvoy2021selfish,MultiAgentLearning}, and introspection. Traditional models of evolutionary game dynamics often assume that players possess complete knowledge and unrestricted access to the full strategy set. However, this assumption does not always hold in real-world contexts, as individuals differ in their knowledge, experience, and cognitive capacity. Moreover, factors such as emotional states and external influences can further shape an individual's strategic choices, leading to heterogeneity in strategies accessible to each individual \cite{Jiang2018}. When players in a game possess different strategy sets, the situation can be described using the concept of the ``hypergame", first introduced by Bennett in 1977 \cite{Bennett1977}. The hypergame framework captures realistic scenarios in which players operate with asymmetric perceptions of the game and unequal access to strategic options.

A fundamental example of strategic adaptation arises in the repeated prisoner's dilemma, where individuals condition their future actions on prior interactions. Strategies such as ``tit-for-tat"\cite{Nowak1992TFT} and ``win-stay, lose-shift"\cite{Nowak1993:WSLS} have been shown to facilitate cooperation in repeated interactions. Classical evolutionary models typically assume symmetry among players, where all individuals have identical strategy sets and can accurately assess their opponents' behaviors. In contrast, the hypergame framework introduces strategy sets' asymmetry—players may differ in their available strategies or in their perception of the game—which fundamentally reshapes the evolutionary dynamics. This deviation from traditional assumptions can give rise to novel equilibria and cooperation patterns that do not emerge under standard symmetric models.

Beyond the classical two-strategy prisoner's dilemma, cooperation emerges in more complex multi-strategy settings \cite{JTBoptionalgames,Hauert2002,scienceEmergenceofCostlyPunishment,WAHL1999:continue_PD,SzabPREthree}, where strategic interactions can lead to cyclic dominance patterns or other emergent behaviors. For instance, in voluntary participation games, opting out of interaction—acting as a ``loner"—can yield higher payoffs than mutual defection, but lower payoffs than mutual cooperation. Such dynamics often result in  oscillatory cycles reminiscent of rock-paper-scissors patterns, where cooperators, defectors, and loners take turns dominating the population. While voluntary participation can reduce exploitation and partially resolve social dilemmas, it does not always ensure the stable establishment or long-term prevalence of cooperation.

We depart from the conventional view that players rely on the same strategy set and that behavioral adaptation takes place on only one time scale, introducing instead a hypergame-based framework\cite{Imhof2007TFTWSLS,Baek2016comparreacandm1}. Specifically, we consider a two-timescale evolutionary process. On the fast timescale, individuals engage in strategic interactions guided by introspection reasoning dynamics \cite{Hauser2019, Couto2022}, which influence their choice of strategies. Initially developed for asymmetric games, introspection reasoning allows players to evaluate their strategic options and adjust accordingly. On the slow timescale, the composition of strategy sets evolves based on players' relative success. Similar timescale separation arguments have been employed in studies of reciprocal behavior in human societies \cite{McAvoy2021selfish,Schmid2022}.

In this study, we develop minimal, prototypical models that capture the fundamental features of hypergame dynamics and reveal their robust behavioral patterns. Our primary focus is a three-strategy system exhibiting the rock-paper-scissors relationship, which has been extensively studied in evolutionary systems. We examine two distinct scenarios. In the first, each player selects two out of the three available strategies—ensuring uniform set size but asymmetric composition. In the second, players have access to any combination of strategies. By applying the hypergame framework to both well-mixed and spatially structured populations, we analyze how prevailing strategy sets depend on player rationality and initial conditions. Our findings demonstrate that higher levels of player rationality facilitate cooperation, ultimately enhancing the overall performance of the system.

\section{Preliminaries and problem formulation}  
\begin{figure*}[htbp]
	\centering
	\includegraphics[width=0.8\linewidth]{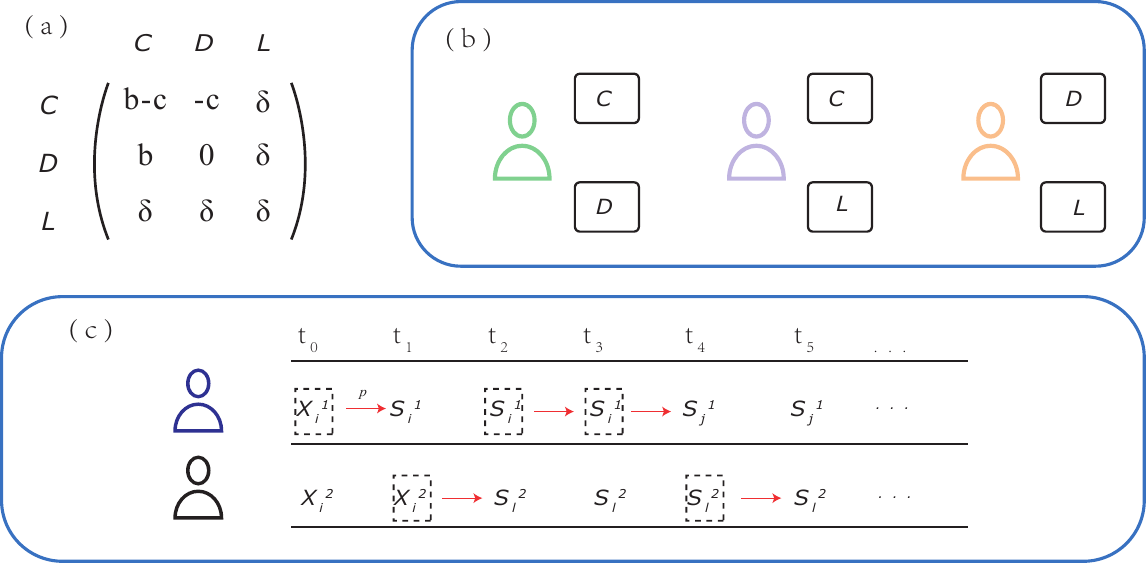}
	\caption{
		(a) The prisoner's dilemma game with voluntary participation. Cooperators provide a benefit $b$ to the other player at a cost $c$, while defectors contribute nothing. In cooperation interactions (CC), players gain $b-c$, while in mutual defection (DD), the gains are $0$. If one player cooperates and the other defects, the cooperator gains $-c$, and the defector gains $b$. If a player opts out of the game (L), the game is invalid, and both players receive a fixed loner payoff ($\delta$). (b)Due to inherent or externally imposed inequalities, players may have access to different subsets of the full strategy set. When constrained to two strategies, three distinct sets emerge: $\{C, D\}$, $\{C, L\}$, and $\{D, L\}$. Without constraints, there are seven possible sets: $\{C\}$, $\{D\}$, $\{L\}$, $\{C, D\}$, $\{C, L\}$, $\{D, L\}$, and $\{C, D, L\}$. (c)  Schematic illustration of introspection dynamics in repeated interactions. Two players, each with potentially different strategy sets, adapt their choices through introspection-based learning.  Initially, each player randomly selects a strategy from their set (e.g., player 1 chooses $X_i^1$ and player 2 chooses $X_i^2$), yielding a payoff $\pi(X_i^1, X_i^2)$ to player 1.  In each learning step, one player is randomly selected to revise their strategy. If player 1 is chosen, he randomly draw a new candidate strategy $S_i^1$ from his own set and compare the hypothetical payoff $\pi(S_i^1, X_i^2)$ to the current payoff $\pi(X_i^1, X_i^2)$. The strategy is adopted with a probability given by the Fermi function: $1 / (1 + \exp(-w \Delta \pi))$, where $\Delta \pi = \pi(S_i^1, X_i^2) - \pi(X_i^1, X_i^2)$ and $w$ denotes the introspection strength. This process repeats iteratively, driving behavioral adaptation over time.}
	\label{fig:1}
\end{figure*}

To investigate evolutionary dynamics within the hypergame framework, we extend the classic prisoner's dilemma to include three strategic options (Fig. \ref{fig:1}(a)): Cooperation ($C$), Defection ($D$), and Loner ($L$). Cooperators and defectors willingly engage in a game involving risk and potential reward, while loners opt out and receive a fixed payoff \(\delta\).  When a player encounters a loner, no game takes place; instead, the player is compelled to act as a loner as well, receiving the same fixed payoff \(\delta\). In a well-mixed population, standard replicator dynamics reveal that these three strategies exhibit cyclic dominance, closely resembling the well-known rock-paper-scissors interactions\cite{Hauert2002}.  

In the hypergame framework, each player is limited to a subset of the original strategy set, capturing the realistic notion that decision-making is often constrained by heterogeneity in knowledge, experience, or accessible alternatives. When applied to the prisoner's dilemma with voluntary participation, this formulation gives rise to seven distinct player types(no strategy in the empty subset and thus unconsidered),  each characterized by a unique subset of strategies. Although the underlying game retains a symmetric payoff structure, strategic asymmetries emerge naturally due to variation in players’ strategy sets. As a result, interactions are shaped not only by the actions selected but also by the strategic lenses through which players interpret the game. This heterogeneity gives rise to complex evolutionary dynamics that cannot be adequately captured by standard symmetric models.

Individual heterogeneity manifests in two primary ways: differences in the size of available strategy sets and variations in their composition. We first examine cases where all players have the same number of strategies but differ in which strategies are included. We then extend the analysis to scenarios where both the number and composition of strategies vary. In this study, the former includes three distinct strategy sets (each containing two strategies, as illustrated in Fig. \ref{fig:1}(b)), while the latter encompasses all seven possible strategy sets.  

To explore the impact of heterogeneity in strategy sets on evolutionary dynamics, we consider two player types, player 1 and player 2, each characterized by a strategy set \( S_h \) (\( h \in \{1,2\} \)). The strategy sets may differ between players, i.e., \( S_1 \neq S_2 \). Over time, players adapt to opponents by dynamically exploring strategies available in their strategy sets. To model these strategic adjustments, we employ introspection dynamics, a method well-suited for asymmetric games.  

Following the introspection dynamics framework\cite{Couto2022} (Fig. \ref{fig:1}(c)), players begin with randomly assigned strategies, \( s_1 \in S_1 \) and \( s_2 \in S_2 \). At each learning step, a player, denoted as player \( h \), is randomly selected to update their strategy. The player attempts to adopt a new strategy \( s_{h}^{\prime} \in S_h \). Let \( \pi_h \) be the payoff associated with the player's current strategy \(s_{h}\) and \( \pi_{h}^{\prime} \) be the payoff from the new strategy \(s_{h}^{\prime}\) while the opponent's strategy remains unchanged. The probability that player \( h \) switches to the new strategy is given by the Fermi function
\begin{equation} \label{Equ:Fermi}
	\rho_w (\pi_{h}^{\prime}-\pi_{h})= \frac{1}{1+e^{-w(\pi_{h}^{\prime}-\pi_{h})}}.  
\end{equation}  
Here, \( w \ge 0 \) represents the introspection strength, which quantifies the extent to which payoffs influence strategic decisions. For small \( w \), selection is weak, meaning strategy adoption is largely stochastic, with payoffs playing a minor role. Conversely, large \( w \) implies strong selection, where players predominantly adopt strategies that yield higher payoffs. Introspection strength thus serves as a proxy for rationality: Lower values correspond to slower adaptation and possible delays in recognizing strategies with higher payoffs, whereas higher values reflect faster shifts toward more beneficial strategies.

Given the current state \( (s_{1}^{i}, s_{2}^{j}) \), the probability of transitioning to a new state \( (s_{1}^{k}, s_{2}^{l}) \) within one time step is defined by the transition matrix, given as

\begin{equation} \label{Eqa:Mij}
	M_{ij,kl}=
	\begin{cases}
		\frac{1}{2(m-1)} \cdot \rho_w\left(\pi(s_{1}^{k},s_{2}^{l})-\pi(s_{1}^{i},s_{2}^{j}) \right), & \text{if } i \neq k, j=l \text{ (Player 1 updates)} \\  
		\frac{1}{2(n-1)} \cdot \rho_w\left(\pi(s_{2}^{l},s_{1}^{k}) -\pi(s_{2}^{j},s_{1}^{i})\right), & \text{if } i=k, j \neq l \text{ (Player 2 updates)} \\  
		0, & \text{if } i \neq k, j \neq l \text{ (Both players update)} \\  
		1-\frac{1}{2} \sum_{p \neq i}^{m} \frac{\rho_w\left(\pi(s_{1}^{p},s_{2}^{j}) - \pi(s_{1}^{i},s_{2}^{j})\right)}{m-1} \\ 
		\quad -\frac{1}{2} \sum_{q \neq j}^{n} \frac{\rho_w\left(\pi(s_{2}^{q},s_{1}^{i}) - \pi(s_{2}^{j},s_{1}^{i})\right)}{n-1}, & \text{if } i=k, j=l \text{ (No update).}
	\end{cases}
\end{equation}

This update rule is iteratively applied over multiple time steps, \( t \in \{1, . . ., T\} \), generating a sequence of payoffs \( \{\pi_{hh^{\prime}}(t)\}_{t=1}^{T} \), where \( \pi_{hh^{\prime}}(t) \) represents player \( h \)'s payoff against player \( h^{\prime} \) at time \( t \), and \( T \) is the total number of updating steps. We define the time-averaged payoff as 

\begin{equation}
	\tilde{\pi}_{S_{h}: S_{h^{\prime}}} =\lim_{T \to \infty } \frac{1}{T} \sum_{t=1}^{T} \pi_{h{h^{\prime}}}(t).
\end{equation}  

This quantity represents the average payoff of a player adopting strategy set \( S_h \) when interacting with an opponent using strategy set \( S_{h^{\prime}} \). The underlying stochastic process unfolds within the set of all possible strategy profiles \( (s_1^1, s_2^1), (s_1^1, s_2^2), ..., (s_1^m, s_2^n) \), where \( m \) and \( n \) denote the number of strategies available to players 1 and 2, respectively. Each state corresponds to the currently adopted strategies of both players. Our objective is to analyze the mathematical properties of this dynamic process.

To derive explicit results, we note that strategy updating depends only on the current state and is independent of the states before. This allows us to model the system as a Markov chain. Given the current state \( (s_1^i, s_2^j) \), the probability of transitioning to \( (s_1^k, s_2^l) \) in one time step follows the transition probabilities defined in Eq. (\ref{Eqa:Mij}). The factor \( 1/2 \) accounts for the random selection of a player to update. The  factors \( 1/(m-1) \) and \( 1/(n-1) \) represent the random selection of an alternative strategy. The function \( \rho_w(\Delta \pi) \) determines the probability of adopting the alternative strategy as given in Eq. (\ref{Equ:Fermi}). These transition probabilities are assembled into an \( mn \times mn \) matrix, denoted as \( M = (M_{ij,kl}) \). Here, the first pair of indices represents the previous state \( (s_1^i, s_2^j) \), while the second pair corresponds to the new state \( (s_1^k, s_2^l) \). The matrix \( M \) is nonnegative and row-stochastic.  

Using \( M \), we characterize the probability distribution over all states at time \( t \), given an initial distribution. Let \( v_{ij}(t) \) denote the probability that the process is in state \( (s_1^i, s_2^j) \) at time \( t \). The vector (\( v_{i,j}(t) \)), along with the corresponding payoff vectors \( \pi_{v1} \) and \( \pi_{v2} \), compiles the probabilities and payoffs for both players across states. Assuming a finite introspection strength \( w \), the transition matrix \( M \) is primitive, ensuring that (\( v_{i,j}(t) \)) converges to a unique stationary distribution \( \tilde{v} \), independent of the initial conditions. The stationary distribution satisfies the eigenvector equation,
\begin{equation} \label{v=vm}
	\tilde{v} =\tilde{v} M,  
\end{equation}  
as dictated by the Perron–Frobenius theorem\cite{meyer2023matrix}. In equilibrium, the expected payoffs for players 1 and 2 are given by

\begin{align}
	\tilde{\pi}_{S_{1}:S_{2}} & = \frac{\tilde{v}\cdot \pi_{v1} }{\tilde{v}\cdot 1},  
	\label{calcu:pi1} \\
	\tilde{\pi}_{S_{2}: S_{1}} & = \frac{\tilde{v}\cdot \pi_{v2} }{\tilde{v}\cdot 1},  
	\label{calcu:pi2}
\end{align}  
where \( \mathbf{1} \) denotes a vector of the same dimension as \(\pi_{v1}\) or \(\pi_{v2}\), with all entries equal to 1. The denominators normalize \( \tilde{v} \) so that its elements sum to one, ensuring it represents a proper probability distribution.

\section{Main results}
\subsection{Tournament competition}
Our investigation begins with a series of tournaments, analyzing pairwise competitions among players employing different strategy sets. For each possible pair of strategy sets, $S_i$ and $S_j$, we evaluate their relative performance by computing the average payoffs (Eqs.~\ref{calcu:pi1} and \ref{calcu:pi2}) obtained when two players—each restricted to their respective strategy sets—interact in pairwise competition. Performance evaluation encompasses both direct competition between different strategy sets and self-play scenarios, where a player's strategy set is tested against itself. 

We begin by examining the case in which each strategy set has a fixed size of two, i.e., $\left| S_i \right| = 2$, under weak introspection strength. Pairwise competitions across different benefit-to-cost ratios $b/c$ reveal a consistent pattern: Strategy sets containing cooperation are systematically outcompeted by those that contain defection (e.g., $\tilde{\pi} _{CL: CD} < \tilde{\pi} _{CD:CL}$). This outcome is intuitive—under weak introspection strength, players update their strategies in a largely stochastic manner, reducing the impact of payoffs on decision-making. Consequently, cooperation becomes a disadvantageous trait, defection gains an advantage, and opting out (loner) yields neutral outcomes. Notably, in self-play evaluations, the strategy set $\{C, D\}$ consistently dominates, except when the cooperation benefit-to-cost ratio is exceptionally high ($b=1.5$). This is primarily due to the relatively high payoff associated with opting out of the game when $b=1.5$.

Next, we systematically explore the performance of different strategy sets under moderate ($w=1$) and strong ($w=10$) introspection strengths across various benefit-to-cost ratios (see Table \ref{tab:1}). Although ${C, L}$ continues to underperform in direct competition, the competitive advantage of ${C, D}$ and ${D, L}$ over ${C, L}$ diminishes as introspection strength increases. The underlying mechanism is clear: As players become more adaptive,  $\{C, L\}$ players can effectively recognize exploitation and swiftly shift to the loner strategy ($L$) for self-protection. Meanwhile, in self-play scenarios, $\{C, L\}$ surpasses $\{C, D\}$ to become the dominant strategy set. This shift is attributed to the increased adaptive capacity, which enhances $\{C, L\}$’s ability to sustain mutual cooperation through self-reciprocity.

To provide a holistic measure of strategy set effectiveness, we compute a combined score (see Table \ref{tab:2}), representing the sum of a strategy set’s average payoffs across all three possible competitor strategy sets, including itself. The performance of each strategy set shows a strong dependence on both the benefit-to-cost ratio and the level of introspection strength. Specifically, under weak introspection ($w=0.1$), $\{D, L\}$ prevails at low $b/c$ values, while $\{C, D\}$ prevails as $b/c$ increases. Under moderate introspection strength ($w=1$), $\{D, L\}$ initially dominates for small $b/c$, but as $b/c$ increases, $\{C, L\}$ emerges as the superior strategy set. For strong introspection ($w=10$), $\{D, L\}$ retains dominance at low $b/c$, while $\{C, L\}$ becomes the optimal strategy as $b/c$ rises.

Let us now consider the tournament competition between the seven possible strategy sets: $\{C\}$, $\{D\}$, $\{L\}$, $\{C, D\}$, $\{C, L\}$, $\{D, L\}$, and $\{C, D, L\}$ (see Table \ref{tab:3}). Players with single-strategy set have no room to adapt in the repeated game. For example, the player with the strategy set $\{C\}$ unconditionally cooperates in repeated interactions. Under weak introspection strength, the single-strategy set $\{L\}$ emerges as the most successful, indicating that abstaining from the game yields higher payoffs in this regime. As introspection strength increases, $\{D, L\}$ takes the lead at small $b/c$ values, while $\{C, L\}$ becomes dominant once $b/c$ exceeds a critical threshold. Notably, under strong introspection, the self-reciprocal nature of $\{C, L\}$ further enhances its competitiveness.

\subsection{Theoretical analysis on pairwise interaction dynamics}
Our tournament analysis indicates that the strategy set $\{C, L\}$  becomes increasingly dominant as both the introspection strength \( w \) and the benefit-to-cost ratio \( b/c \) rise. To account for this observation, we examine the introspection dynamics about the strategy set $\{C, L\}$. According to the tournament results, the key factor underlying its success lies in its ability to generate favorable outcomes when $\{C, L\}$ players interact with each other. For an interaction between two $\{C, L\}$ players, the resulting subgame matrix is 
\[
\begin{array}{cccc}
	& & C & L \\
	& C & (b-c, b-c) & (\delta, \delta) \\
	& L & (\delta, \delta) & (\delta, \delta) \\
\end{array}.
\]
There are four possible states, $(C, C)$, $(C, L)$, $(L, C)$, and $(L, L)$. In this order, the corresponding transition matrix is
\[
M =
\begin{pmatrix}
	1 - \frac{1}{1 + e^{w(b - c - \delta)}} & \frac{1}{2(1 + e^{w(b - c - \delta)})} & \frac{1}{2(1 + e^{w(b - c - \delta)})} & 0 \\
	\frac{1}{2(1 + e^{-w(b - c - \delta)})} & \frac{3}{4} - \frac{1}{2(1 + e^{-w(b - c - \delta)})} & 0 & \frac{1}{4} \\
	\frac{1}{2(1 + e^{-w(b - c - \delta)})} & 0 & \frac{3}{4} - \frac{1}{2(1 + e^{-w(b - c - \delta)})} & \frac{1}{4} \\
	0 & \frac{1}{4} & \frac{1}{4} & \frac{1}{2}
\end{pmatrix}
\]
The stationary distribution $\tilde{v}=(v_{CC}, v_{CL}, v_{LC}, v_{LL} )$ is given by
\begin{align}
	v_{CC} &= \frac{3e^{2w(b - c - \delta)} + e^{w(b - c - \delta)}}{3e^{2w(b - c - \delta)} + 10e^{w(b - c - \delta)} + 3}, \\
	v_{CL} = v_{LC} = v_{LL} &= \frac{3e^{w(b - c - \delta)} + 1}{3e^{2w(b - c - \delta)} + 10e^{w(b - c - \delta)} + 3}.
	\label{eq:vcl}
\end{align} 
Then,
\begin{equation}
	\tilde{\pi}_{CL:CL} = \frac{3e^{2w(b - c - \delta)} + e^{w(b - c - \delta)}}{3e^{2w(b - c - \delta)} + 10e^{w(b - c - \delta)} + 3}(b - c -\delta) +\delta.
	\label{eq:pi_self}
\end{equation}
Since Eq\(.~\eqref{eq:vcl}\) is a decreasing function of \( w \), it follows that \(\tilde{\pi}_{CL:CL}\) increases monotonically with \( b \) and \( w \). This means that  \( b \) increases the reward for mutual cooperation, and a higher \( w \) leads to a greater probability of remaining in the mutually beneficial state $(C, C)$. Specifically, when \( w = 0 \), the system visits each state with equal probability, i.e., \( v_{CC} = v_{CL} = v_{LC} = v_{LL} = \frac{1}{4} \). In the limit  \( w \to \infty \), we have \( v_{CC} \to 1 \) and hence \( \tilde{\pi}_{CL:CL} \to b - c \), the payoff for mutual cooperation. This means that self-reciprocity between $\{C, L\}$ players reinforces as selection increases.

Achieving success in the tournament requires not only the ability to sustain reciprocity, but also the capacity to resist exploitation by self-interested strategies. In the context of the hypergame, only defectors impose direct costs on cooperators, making the strategy set $\{D\}$ the primary threat. To evaluate the robustness of the $\{C, L\}$ strategy set under such conditions, we examine the dynamics of its interaction with the strategy set $\{D\}$. For such interaction, two states $(C, D)$ and $(L, D)$ are possible, which correspond to cooperation or opting out when facing defection. These states yield the respective payoff pairs $(-c, b)$ and $(\delta, \delta)$. The transitions between these states are governed by the following matrix
\[
M' =
\begin{pmatrix}
	1 - \frac{1}{2(1 + e^{w(-c - \delta)})} & \frac{1}{2(1 + e^{w(-c - \delta)})} \\
	\frac{1}{2(1 + e^{w(c + \delta)})} & 1 - \frac{1}{2(1 + e^{w(c + \delta)})}
\end{pmatrix}
\]
Then, we get the following stationary distribution $\tilde{v}=(v_{CD}, v_{LD})$
\begin{align}
	v_{CD} &= \frac{e^{-w(c + \delta)} + 1}{e^{-w(c + \delta)} + e^{w(c + \delta)} + 2}, \\
	v_{LD} &= \frac{e^{w(c + \delta)} + 1}{e^{-w(c + \delta)} + e^{w(c + \delta)} + 2}.
\end{align}
The expected payoff for a $\{D\}$ player interacting with a  $\{C, L\}$ player is then given by
\begin{equation}
	\tilde{\pi}_{D, CL} = \frac{e^{-w(c + \delta)} + 1}{e^{-w(c + \delta)} + e^{w(c + \delta)} + 2}b + \frac{e^{w(c + \delta)} + 1}{e^{-w(c + \delta)} + e^{w(c + \delta)} + 2}\delta.
\end{equation}
	It can be verified that the payoff $\tilde{\pi}_{D, CL}$ declines monotonically with increasing introspection strength \( w \). As \( w \) rises, individuals with the set $\{C, L\}$  become more adept at avoiding exploitation, thereby diminishing defectors' gains. 
	$\{D\}$ players lose their advantage over $\{C, L\}$ players — that is, $\tilde{\pi}_{CL:CL} \ge \tilde{\pi}_{D,CL}$)— once \( w \) exceeds a critical threshold
	\begin{equation}
		w \ge \frac{x^*}{b-c-\delta},
	\end{equation}
	where \( x^* \) solves
	\[
	e^{(1 + \alpha) x^*} - \alpha e^{x^*} - 3(1 + \alpha) = 0, \quad \text{with} \quad \alpha = \frac{c+\delta}{b-c-\delta}.
	\]
For instance, setting \(\delta=0.25\) and \(c=1\),the critical thresholds are \(w \ge 1.56\) for \(b=1.7\), \(w \ge 1.56\) for \(b=1.7\), and \(w \ge 1.27\) for \(b=1.9\). These results demonstrate that stronger introspection more readily eschews the threat of exploitation.

The combination of strategies \(L\) and \(C\) can outperform unconditional defectors once the \(w\) exceeds a certain threshold. This indicates that the presence of \(L\) helps to stabilize cooperation against pure defection. However, it remains unclear whether the addition of \(L\) to defectors, forming the set \(\{D, L\}\), would further enhance their competitive advantage. To clarify this point, we examine the threshold condition under which \(\{C, L\}\) can dominate \(\{D, L\}\), namely \(\tilde{\pi}_{CL:CL} \geq \tilde{\pi}_{DL:CL}\).  

For the interaction between a \(\{C, L\}\) player and a \(\{D, L\}\) player, the resulting subgame is represented by the payoff matrix  
\[
\begin{array}{cccc}
	& & D & L \\ 
	& C & (-c, b) & (\delta, \delta) \\ 
	& L & (\delta, \delta) & (\delta, \delta) \\
\end{array}.
\]  

This subgame allows four possible states, \((C, C)\), \((C, L)\), \((L, C)\), and \((L, L)\). In this order, the corresponding transition matrix is  
\[
M =
\begin{pmatrix}
	1 - \frac{1}{2(1 + e^{w(b - \delta)})}-\frac{1}{2(1 + e^{w(- c - \delta)})} & \frac{1}{2(1 + e^{w(b - \delta)})} & \frac{1}{2(1 + e^{w(- c - \delta)})} & 0 \\
	\frac{1}{2(1 + e^{-w(\delta-b)})} & \frac{3}{4} - \frac{1}{2(1 + e^{-w(\delta-b)})} & 0 & \frac{1}{4} \\
	\frac{1}{2(1 + e^{-w(b + \delta)})} & 0 & \frac{3}{4} - \frac{1}{2(1 + e^{-w(b + \delta)})} & \frac{1}{4} \\
	0 & \frac{1}{4} & \frac{1}{4} & \frac{1}{2}
\end{pmatrix}.
\]  

The stationary distribution \(\tilde{v}=(v_{CD}, v_{CL}, v_{LD}, v_{LL} )\) is given by  
\begin{align}
	v_{CD} &= \frac{e^{w( - c - \delta)} + e^{w(b - \delta)}+6e^{w(b- c - 2\delta)}}{10e^{w( - c - \delta)} + 10e^{w(b - \delta)} + 6e^{w(b- c - 2\delta)}+6}, \\
	v_{CL}  &= \frac{5e^{w( - c - \delta)} + e^{w(b - \delta)}+2}{10e^{w( - c - \delta)} + 10e^{w(b - \delta)} + 6e^{w(b- c - 2\delta)}+6},\\
	v_{LD}  &= \frac{e^{w( - c - \delta)} + 5e^{w(b - \delta)}+2}{10e^{w( - c - \delta)} + 10e^{w(b - \delta)} + 6e^{w(b- c - 2\delta)}+6},\\
	v_{LL}  &= \frac{3e^{w( - c - \delta)} + 3e^{w(b - \delta)}+2}{10e^{w( - c - \delta)} + 10e^{w(b - \delta)} + 6e^{w(b- c - 2\delta)}+6}.
\end{align}  

Accordingly, the expected payoff of \(\{D, L\}\) against \(\{C, L\}\) is  
\begin{equation}
	\tilde{\pi}_{CD:CL} =\frac{e^{w( - c - \delta)} + e^{w(b - \delta)}+6e^{w(b- c - 2\delta)}}{10e^{w( - c - \delta)} + 10e^{w(b - \delta)} + 6e^{w(b- c - 2\delta)}+6}\cdot b + \frac{9e^{w( - c - \delta)} + 9e^{w(b - \delta)}}{10e^{w( - c - \delta)} + 10e^{w(b - \delta)} + 6e^{w(b- c - 2\delta)}+6}\cdot \delta.
\end{equation}  

Our analysis shows that when the benefit \(b\) is relatively small, \(\{D, L\}\) tends to outperform \(\{C, L\}\) more easily than unconditional defectors, implying that the threshold for \(\tilde{\pi}_{CL:CL} \geq \tilde{\pi}_{DL:CL}\) becomes higher. For instance, setting \(\delta=0.25\) and \(c=1\), the critical thresholds are \(w \geq 4.53\) for \(b=1.5\), \(w \geq 1.97\) for \(b=1.7\), and \(w \geq 1.37\) for \(b=1.9\).  Despite the fact that the combination of \(L\) and \(D\) can enhance the competitive advantage of defectors in some cases, the combination of \(\{C, L\}\) always maintains its superiority when the parameter \(w\) is sufficiently large. This indicates that as \(w\) increases, \(\{C, L\}\) becomes more robust, ensuring stable cooperation even in the presence of defection strategies.

Taken together, these results show that increasing introspection simultaneously enhances the self-reciprocal nature of cooperation within $\{C, L\}$ and its defensive capacity against defection. These theoretical insights align closely with our simulation findings and offer further support for the analysis of evolutionary dynamics.

\subsection{Evolutionary dynamics of strategy sets}  
Building upon the tournament competitions between diverse strategy sets, we now explore the evolutionary dynamics of these strategy sets. To do this, we consider a large, well-mixed population of players, with strategy evolution occurring on two distinct time scales. 

In the fast term, players in the population adhere to fixed strategy sets, with $x_S$ denoting the fraction of players adopting strategy set $S$. These players are randomly paired to engage in pairwise interactions, following the approach outlined in the previous section. When player $S_i$ and player $S_j$ interact, their resulting payoffs are given by $\tilde{\pi}_{S_1: S_2}$, as defined in Eq. \ref{calcu:pi1}. 

In the slow term, the composition of the population evolves as players with different strategy sets compete to survive. Various mathematical frameworks can describe the evolution of strategy sets; here, we adopt the replicator equation\cite{Cressman2014_replicatodynamic}  to model these slow dynamics. In particular, when one strategy set consistently yields higher payoffs than another, the inferior strategy set is eventually eliminated—a fundamental property that holds under any ``payoff-monotone'' dynamics.

\begin{figure*}[!htbp]
	\centering
	\includegraphics[width=0.8\linewidth]{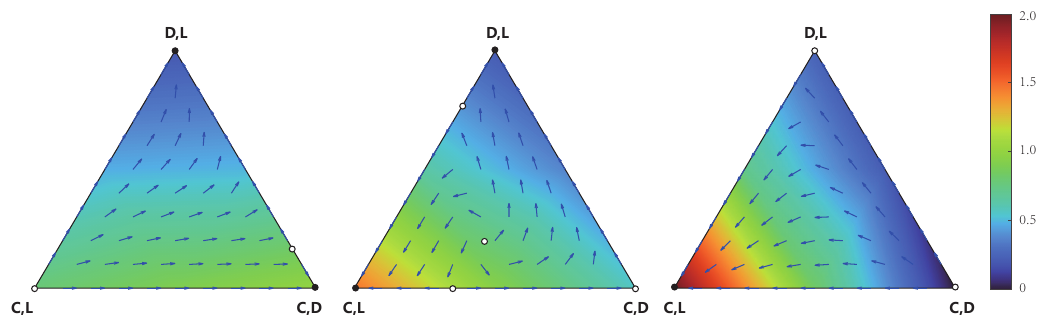}
	\caption{Evolutionary dynamics of three strategy sets. To investigate the evolution of diverse strategy sets, replicator dynamics are employed. Population members can adopt one of three strategy sets: $\{ C, D\}$, $\{ C, L\}$, or $\{ D, L\}$. The simplex's three vertices symbolize homogeneous populations for each of the three distinct strategy sets. This depiction illustrates the replication dynamics: $\dot{x}_{S} =x_{S} \left ( P_{S} - \bar{P} \right )$, where $P_{S}$ denotes the payoff of the strategy set $S$ in the population and $\bar{P}$ denotes the average payoff of the population. For populations with proportions $x_{{C, D}}$, $x_{{C, L}}$, and $x_{{D, L}}$ of each strategy set, $P_{\left \{ C,D \right \} } =x_{\left \{ C,D \right \} }\cdot \tilde{\pi} _{CD:CD}+x_{\left \{ C,L \right \} }\cdot \tilde{\pi} _{CD:CL} +x_{\left \{ D,L \right \} }\cdot \tilde{\pi} _{CD:DL}$, and the average payoff $\bar{P}= x_{\left \{ C,D \right \} }\cdot P_{\left \{ C,D \right \} }+x_{\left \{ C,L \right \} }\cdot P_{\left \{ C,L \right \} } +x_{\left \{ D, L \right \} }\cdot P_{\left \{ D, L \right \} }$. The proportion of population members with a particular strategy set changes over time based on whether players with that strategy set achieve an expected payoff above the average. Specifically, with $b = 3$, we explore the dynamics of the three different strategy sets across varying introspection strengths. Under weak introspection strength $w=0.1$, the majority of regions converge to a homogeneous population of $\{D, L\}$, with a small fraction converging to $\{C, D\}$. For moderate introspection strength $w=1$, the population may evolve to different homogeneous populations, contingent on the initial state. Under strong introspection strength $w=10$, $\{C, L\}$ demonstrates near-global stability, converging to a monomorphic population of $\{C, L\}$ for almost all initial states. In addition, the color on the simplex indicates the average gain of the population under the corresponding population composition. The introspection strength varies from left to right: $w=0.1$, $w=1$, $w=10$.}
	\label{fig:2}
\end{figure*}

\begin{figure*}[ht]
	\centering
	\includegraphics[width=0.9\linewidth]{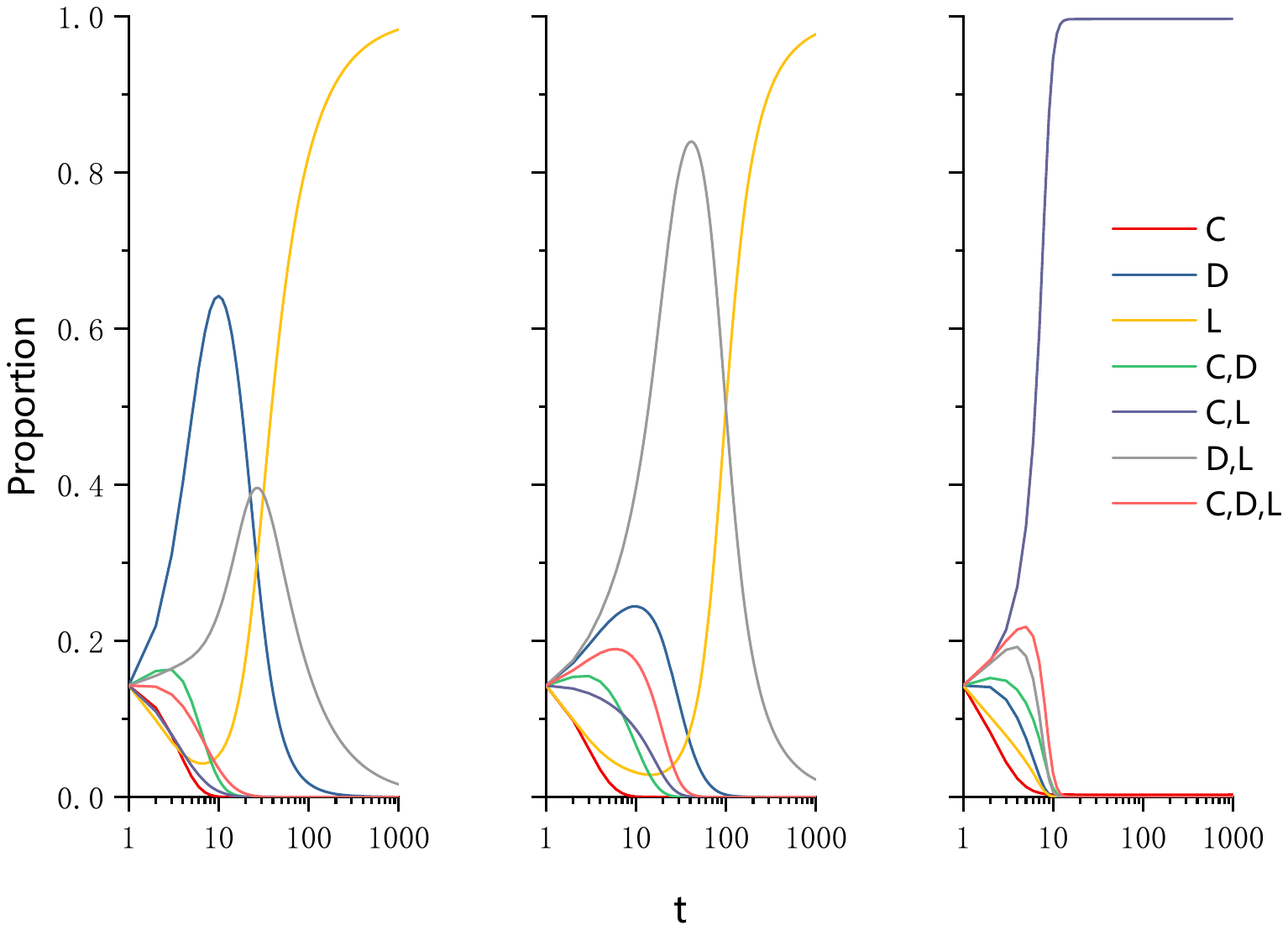}
	\caption{ Evolutionary dynamics of seven strategy sets. To examine the changing frequency of strategy sets in the player population, we employ a simulation approach. During each simulation iteration, player payoffs are calculated following the method illustrated in Fig.\ref{fig:2}. Subsequently, players update their strategies based on the obtained payoffs, with the frequency of strategy set $S$ in the next generation represented as $x_{S }(t+1) =x_{S}(t) \cdot \left ( \frac{P_{S} }{\bar{P}} \right )$. This iterative updating process repeats over numerous generations. The simulation parameters are set as follows: $b=3$, $c=1$, and introspection strength varies from left to right: $w=0.1$, $w=1$, $w=10$. }
	\label{fig:3}
\end{figure*}

To systematically explore the evolutionary dynamics of diverse strategy sets, we separately analyze cases involving three-strategy sets (Fig. \ref{fig:2}) and all seven possible strategy sets (Fig. \ref{fig:3}) across three levels of introspection strength. When only three strategy sets are considered, under weak introspection ($w=0.1$), the population predominantly converges to a low-payoff equilibrium dominated by $\{D, L\}$ players, as defection offers a short-term advantage. However, as introspection strength increases, the $\{C, L\}$ strategy set gains an evolutionary advantage, bolstered by its ability to foster reciprocity and resist exploitation. At high introspection levels, the population consistently converges to a high-payoff equilibrium dominated by $\{C, L\}$ players.  

For the evolutionary race involving seven strategy sets, weak introspection strength favors the strategy set $\{L\}$ players, suggesting that abstaining from participation yields higher payoffs. As introspection strength increases, the population transitions toward dominance by $\{C, L\}$ players, indicating that engagement in the game—coupled with cooperation—becomes the optimal strategy set.  

Thus far, we have mainly focused on the evolutionary dynamics of the well-mixed population. We now turn to a more realistic scenario by examining dynamics on structured populations. Specifically, we consider a population arranged on a two-dimensional lattice with periodic boundary conditions, where each individual occupies a node and interacts with its four nearest neighbors (i.e., von Neumann neighborhood).

After accumulating payoffs from interactions with neighbors, players update their strategy sets asynchronously.  At each evolutionary step, a focal player \(X\) is selected at random and compares his payoff \(E_X\) to that of a randomly chosen neighbor \(Y\), whose payoff is \(E_Y\). Player \(X\) then adopts \(Y\)’s strategy set with probability  
\begin{equation}  
	T = \frac{1}{1+e^{-(E_Y - E_X)/K}},  
\end{equation}
where $K$ introduces stochasticity to account for occasional irrational decisions. \(E_X\) and \(E_Y\) denote the average payoffs of players \(X\) and \(Y\), calculated over their respective neighbor interactions. In the limit $K \to 0$, player X adopts Y’s strategy set deterministically if $E_Y > E_X$. A finite $K$ allows for occasional deviations from strict payoff-based decision-making. Our analysis primarily considers cases with $K = 0.01$.

\begin{figure*}[ht]
	\centering
	\includegraphics[width=0.9\linewidth]{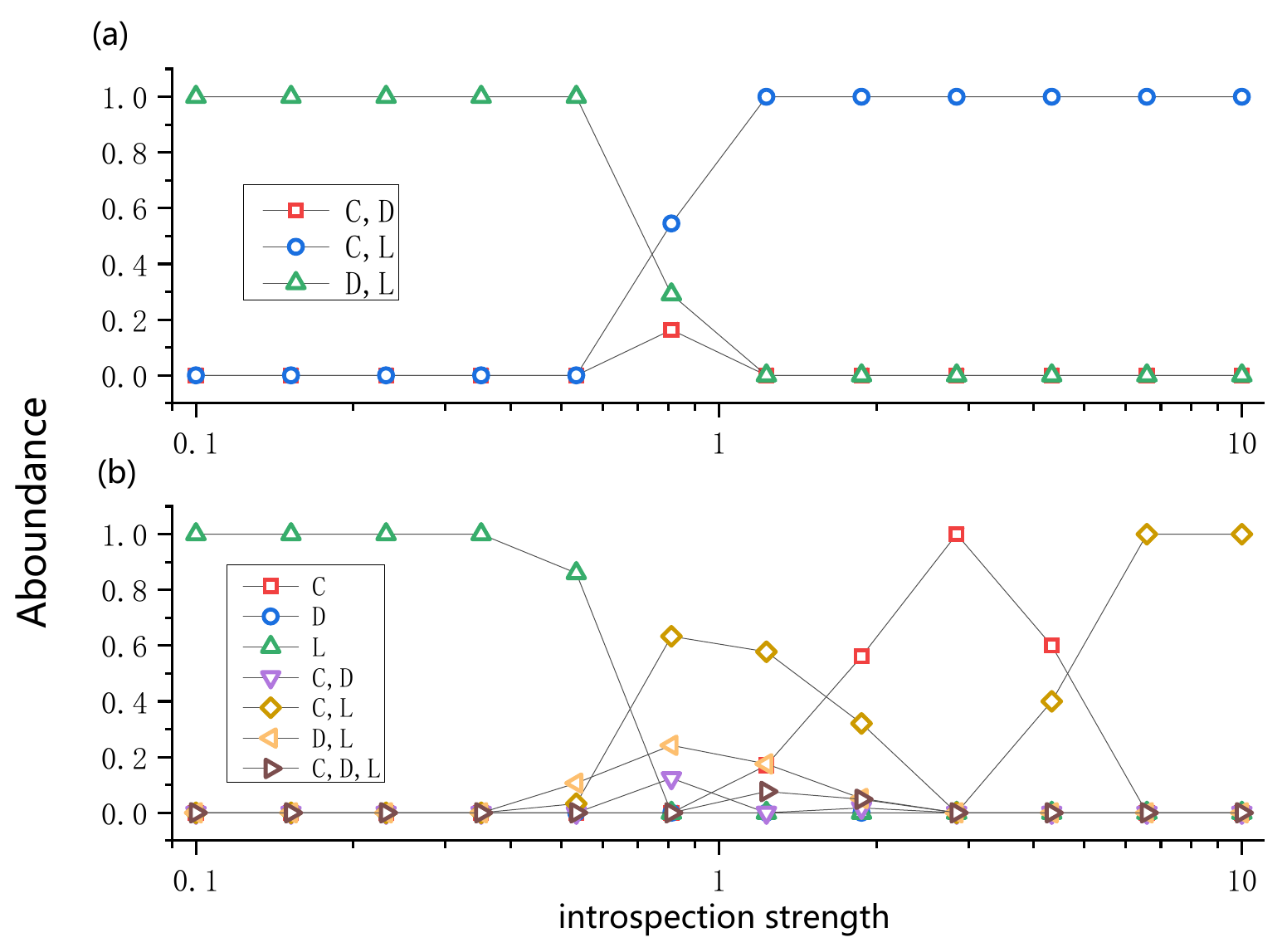}
    \caption{
	 Evolutionary dynamics of the hypergame on a spatially structured population. We consider a $100\times 100$ lattice network with periodic boundary conditions. Each player occupies a node and interacts exclusively with its four nearest neighbors. During each update, a player randomly selects one neighbor for comparison, with the probability of adopting the neighbor's strategy set determined by the relative payoffs. The figure illustrates how the distribution of strategy sets evolves as a function of introspection strength. We examine which strategy sets ultimately dominate the population under varying levels of introspection. Our results show that different introspection strengths lead to qualitatively distinct evolutionary outcomes, with different strategy sets prevailing across the population. Each data point represents the average of 10 independent simulations, each lasting $10^7$ update rounds. Parameters: $b=3$, $c=1$, $K=0.1$.
     }	\label{fig:4}
\end{figure*}

\begin{figure*}[htbp]
	\centering
	\includegraphics[width=0.8\linewidth]{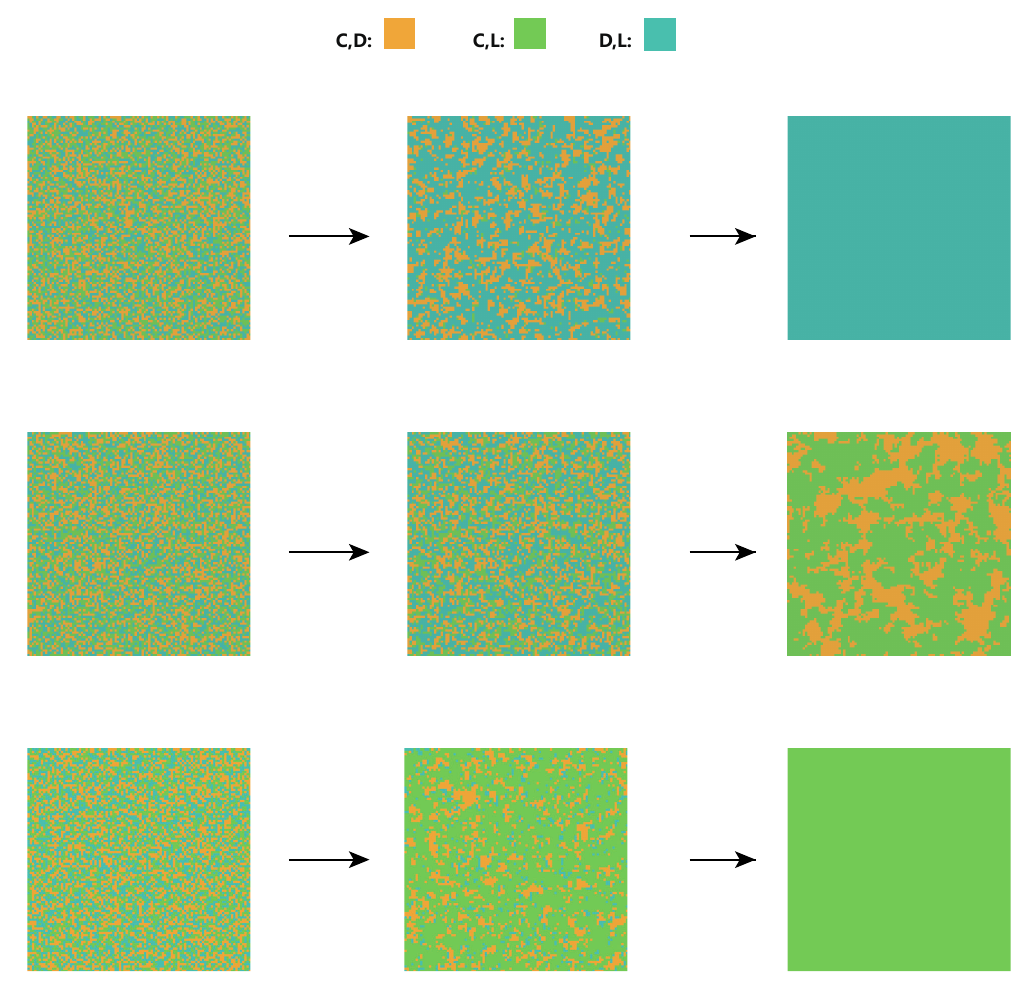}
	\caption{
		Illustrative snapshots of hypergame dynamics with three strategy sets on a square lattice. Fig.~\ref{fig:4} presents three distinct dynamic regimes, and the corresponding snapshots are shown here for visual reference. The three rows correspond to different levels of introspection strength: $w=0.1$, $w=1$, and $w=10$, from top to bottom. These snapshots provide an intuitive visualization of how varying introspection strength influences the spatial distribution of strategy sets.}	\label{fig:5}
\end{figure*}

We systematically investigate how introspection strength influences the evolutionary dynamics in structured populations. When only the three strategy sets $\{C, L\}$, $\{D,L\}$ and $\{C, D\}$ take part in the evolutionary race, the $\{D, L\}$ strategy set dominates(Fig. \ref{fig:4}(a)) for weak introspection strength ($w < 0.6$). However, as introspection strength increases, $\{C, L\}$ gradually displaces $\{D, L\}$ and takes over the population. Further examination of the spatial dynamics (Fig.\ref{fig:5}) reveals that under weak introspection, $\{D, L\}$ expands at the expense of other strategy sets, leading to a monomorphic population. In contrast, as introspection strength increases, $\{C, L\}$ players outcompete $\{D, L\}$ players, ultimately becoming the dominant strategy set.  

When the seven strategy sets take part in the evolutionary race, the dynamics exhibit distinct evolutionary transitions (Fig. \ref{fig:4}(b)). When introspection strength is low ($w<0.5$), the  strategy set $\{L\}$ dominates the population. As introspection strength increases within the range of $0.5<w<2.0$, a coexistence of various strategy sets emerges, with $\{C, L\}$ gaining prominence. For $2<w<5$, the  strategy set $\{C\}$ increasingly prevails, and at certain points (e.g., $w=3.0$), the population converges to a monomorphic state of $\{C\}$ players. At higher introspection levels ($w>5$), $\{C, L\}$ ultimately takes over the population.  

\begin{figure*}[htbp]
	\centering
	\includegraphics[width=0.9\linewidth]{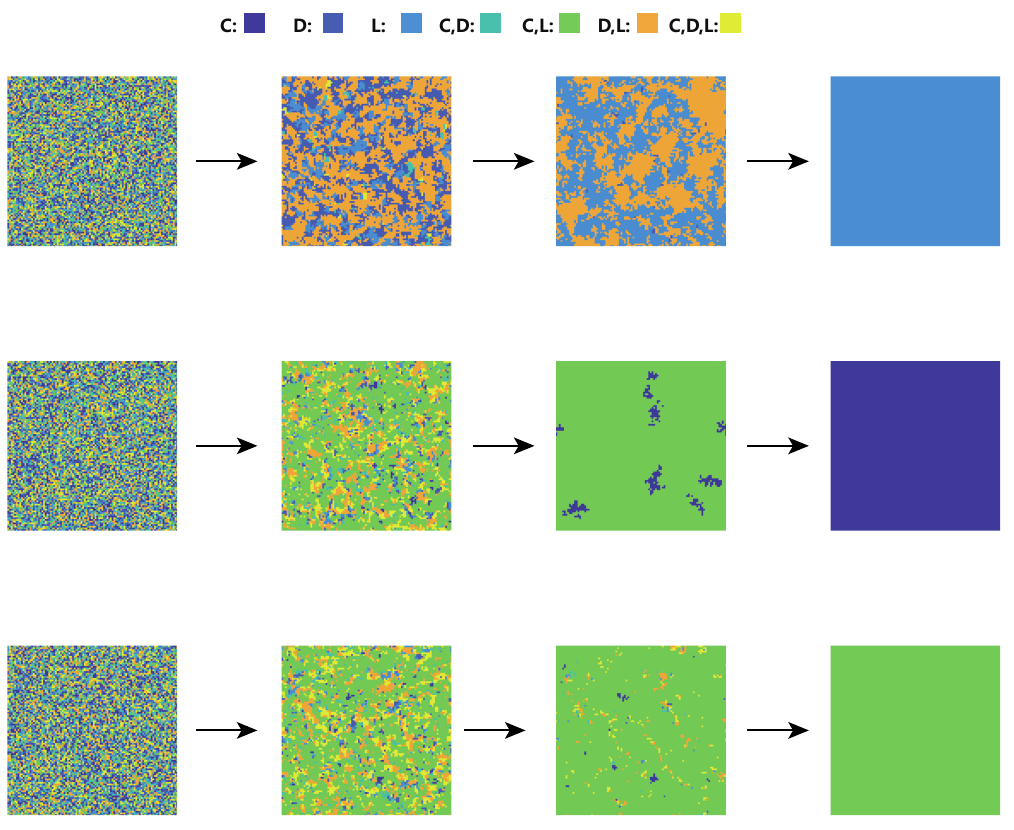}
    \caption{
	Representative snapshots of hypergame dynamics with seven strategy sets on a square lattice. Following the format of Fig.~\ref{fig:5}, three sets of snapshots are provided to aid intuitive understanding. Each row corresponds to a different introspection strength: from top to bottom, $w=0.1$, $w=3$, and $w=10$.}\label{fig:6}
\end{figure*}

Fig. \ref{fig:6} further illustrates these transitions. Under weak introspection, $\{D, L\}$ and $\{L\}$ players gradually assimilate other strategy sets, with $\{D, L\}$ initially gaining an advantage. However, when only $\{D, L\}$ and $\{L\}$ remain, the latter prevails, leading to a monomorphic $\{L\}$ population. As introspection strength increases, $\{C, L\}$ players gain a competitive edge and spread throughout the population. Despite the vulnerability of the $\{C\}$ strategy set to exploitation, it persists within $\{C, L\}$ clusters. In cases where only $\{C\}$ and $\{C, L\}$ remain, $\{C\}$ clusters achieve higher payoffs, allowing them to dominate the population. Further increases in introspection strength ultimately lead to complete dominance by $\{C, L\}$. 

\section{Conclusion}
We introduce an evolutionary hypergame dynamics framework to more accurately capture realistic scenarios in which individuals do not possess full access to the entire strategy space\cite{Inequalityscience,InequalityPNAS}; instead, players operate within heterogeneous strategy sets, reflecting limitations in knowledge, experience, or perception. Interactions between individuals are governed by introspection reasoning, through which players evaluate strategic possibilities in repeated encounters. Based on the outcomes of these introspective interactions, individuals iteratively refine their strategy sets over the course of the evolutionary process. In this study, we investigate the evolutionary implications of this framework within the context of the prisoner's dilemma with voluntary participation. As both the game parameters and introspection strength vary, the system exhibits rich and dynamic behaviors that were not captured by traditional strategy-update-driven evolutionary models.

Evolutionary hypergame dynamics highlight the critical role of heterogeneity in shaping the evolutionary trajectories of strategy sets. Unlike classical evolutionary game theory—which typically assumes uniform strategy access and restricts evolutionary change to the level of strategy selection\cite{PDandSN}—our framework allows strategies and strategy sets to co-evolve. Introspection dynamics, whereby individuals reflect on and assess their own strategic context, serve as a plausible and effective mechanism to model interactions among heterogeneous agents. This reconceptualization of evolutionary change—rooted in the evolution of strategy sets—offers a deeper understanding of how complex social behavior may emerge and stabilize.

Our results reveal that variations in introspection strength fundamentally reshape the evolutionary landscape. Specifically, in the three-strategy prisoner's dilemma, greater introspection strength substantially promotes cooperative behavior. We observe that cooperative behavior becomes increasingly prevalent as the introspection strength increases, regardless of whether the population is well-mixed or spatially structured. In contrast, under weak introspection, strategy sets with defection or the loner strategy tend to prevail.

The model we investigate involves a minimal hypergame setup, in which each individual is assigned a strategy set from the available subsets of a three-strategy space, and interactions occur pairwise. Despite its simplicity, this setting provides a clear description of a class of hypergame dynamics and can be systematically explored Despite its simplicity, this setting captures the core features of hypergame dynamics and can be efficiently explored via simulation. Our approach offers a generalizable framework for analyzing evolutionary processes in settings marked by multiple strategies and heterogeneous agents. It sheds light on how learning, strategic innovation, and adaptation unfold in environments where players operate with incomplete or asymmetric information.

In conclusion, evolutionary hypergame dynamics extend the conceptual reach of evolutionary game theory by incorporating heterogeneity in cognitive access to strategies. This perspective offers a powerful framework for studying the evolution of cooperation, particularly in heterogeneous and complex populations. By allowing both behaviors and strategy sets to co-evolve, our model opens new directions for theoretical and empirical investigations into the foundations of social behavior.

\bibliography{references}   

\clearpage

\clearpage
\begin{table*}[htbp]
	\centering
	\caption{  Self-scoring competitions and pairwise competitions involve players with distinct strategy spaces—$\{C, D\}$, $\{C, L\}$, or $\{D, L\}$. Within their respective strategy spaces, players adapt their strategies to opponents through introspection reasoning. The average payoff (i.e. scoring) for each combination of strategy spaces is computed using Eqs.\ref{calcu:pi1} and \ref{calcu:pi2}, and winning players are identified with underlining.  Parameters: $c=1$, $\delta=0.25$, and $w=0.1$ }
	\begin{tabular}{|c|c|c c|c c|c c|c c c|}
		\hline
		$w$ & $b$ & \multicolumn{2}{|c|}{$\left \{ C,D\right \} :\left \{ C,L \right \}$} & \multicolumn{2}{|c|}{$\left \{ C,D\right \} :\left \{ D,L \right \} $} & \multicolumn{2}{|c|}{$\left \{ C,L\right \} :\left \{ D,L \right \} $} &  \multicolumn{3}{|c|} {self}\\
		\hline
		& & $\tilde{\pi} _{CD:CL} $ & $\tilde{\pi} _{CL:CD} $ & $\tilde{\pi} _{CD:DL} $ & $\tilde{\pi} _{DL:CD} $ & $\tilde{\pi} _{CL:DL}$ & $\tilde{\pi} _{DL:CL} $ & $\tilde{\pi} _{CD:CD} $ & $\tilde{\pi} _{CL:CL} $& $\tilde{\pi} _{DL:DL} $ \\
		\cline{2-11}
		\multirow{5}{*}{0.1} & 1.5 &  \underline{0.6197} & 0.0000 & -0.1290 & \underline{0.4985} & -0.0616 & \underline{0.5616} & 0.2375 & \underline{0.3137} & 0.1887 \\
		\cline{2-11}
		\multirow{5}{*}{} & 2.0 &    \underline{0.8694} & 0.1208 & -0.1349 & \underline{0.6314} & -0.0670 & \underline{ 0.6939} & \underline{0.4750} & 0.4482 & 0.1887 \\
		\cline{2-11}
		\multirow{5}{*}{} & 3.0 & \underline{1.3870}  & 0.3759 & -0.1467 & \underline{0.9101} & -0.0779 & \underline{0.9713} & \underline{0.9500} & 0.7474 & 0.1887 \\
		\cline{2-11}
		\multirow{5}{*}{} & 4.0 & \underline{1.9285} & 0.6485 & -0.1583 & \underline{1.2060} & -0.0885  & \underline{1.2656 } & \underline{1.4251} & 1.0887 & 0.1887 \\
		\cline{2-11}
		\multirow{5}{*}{} & 5.0 & \underline{2.4933} & 0.9382 & -0.1697 & \underline{1.5183} & -0.0990 & \underline{1.5761} & \underline{1.9001} & 1.4747 & 0.1887 \\
		\hline
		
		\multirow{5}{*}{1} & 1.5 &    \underline{0.5390} & 0.0439 & -0.1103 & \underline{0.4384} & 0.0044 & \underline{0.4956}  & 0.1345 & \underline{0.3249} & 0.1985 \\
		\cline{2-11}
		\multirow{5}{*}{} & 2.0 &    \underline{0.7712} & 0.1254 & -0.1313 & \underline{0.5713} & -0.0110 & \underline{0.6154}  & 0.2689  & \underline{0.5603} & 0.1985 \\
		\cline{2-11}
		\multirow{5}{*}{} & 3.0 & \underline{1.3102}  & 0.3533 & -0.1553 & \underline{0.8486} &  -0.0285 & \underline{0.8627}  & 0.5379 & \underline{1.4003} & 0.1985 \\
		\cline{2-11}
		\multirow{5}{*}{} & 4.0 & \underline{1.8721} & 0.6182  & -0.1651 & \underline{1.1241} & -0.0357 & \underline{1.1070 }  & 0.8068 & \underline{2.5575} & 0.1985\\
		\cline{2-11}
		\multirow{5}{*}{} & 5.0 & \underline{2.4211} & 0.8879 & -0.1689  & \underline{1.3940} & -0.0384 & \underline{1.3459}  & 1.0758  & \underline{3.7529} & 0.1985\\
		\hline
		\multirow{5}{*}{10} & 1.5 &    \underline{0.4870} & 0.0920 & 0.0000 & \underline{0.3951} & 0.1250 & \underline{0.3750}  & 0.0000 & \underline{0.4506} &0.2433 \\
		\cline{2-11}
		\multirow{5}{*}{} & 2.0 &    \underline{0.6665} & 0.1667 & 0.0000 & \underline{0.4741} & 0.1250 & \underline{0.4250}  & 0.0000  & \underline{0.9988} & 0.2433 \\
		\cline{2-11}
		\multirow{5}{*}{} & 3.0 & \underline{1.0000}  & 0.3334 & 0.0000 & \underline{0.6321} &  0.1250 & \underline{0.5250}  & 0.0000 & \underline{2.0000} & 0.2433 \\
		\cline{2-11}
		\multirow{5}{*}{} & 4.0 & \underline{1.3334} & 0.5000  & 0.0000 & \underline{0.7902} & 0.1250 & \underline{0.7250 }  & 0.0000 & \underline{3.0000} & 0.2433\\
		\cline{2-11}
		\multirow{5}{*}{} & 5.0 & \underline{1.6667} & 0.6667 & 0.0000  & \underline{0.9482} & 0.1250 & \underline{1.3459}  & 0.0000  & \underline{4.0000} & 0.2433\\
		\hline
	\end{tabular}
	\label{tab:1}
\end{table*}

\clearpage
\begin{table*}[htbp]
	\caption{Tournament winners at various introspection strengths with three strategy sets. Integrated with Table \ref{tab:1}, the combined score for each strategy set aggregates the self-scores and scores from competitions with other distinct strategy sets. Winners were identified at varying $b$-values and introspection strengths $w$, indicated by underlining.}
	\centering
	\begin{tabular}{|c|c|c|c|c|c|c|} 
		\hline
		\multirow{2}{*}{ $w$ }& \multirow{2}{*}{space} &\multicolumn{5}{c|}{\textbf{combined score}}  \\ 
		\cline{3-7} 
		& &$b=1.5$&$b=2$&$b=3$&$b=4$&$b=5$\\
		\hline
		\multirow{3}{*}{0.1} & $\left \{ C,D \right \} $ & 0.7282 &1.2095& \underline{2.1903} & \underline{3.1952} & \underline{4.2237} \\
		& $\left \{ C,L \right \} $ &  0.2520 & 0.5020 & 1.0454 &1.6487 &2.3140 \\
		& $\left \{ D,L \right \} $& \underline{1.2488} & \underline{1.5139} &  2.0701 &2.6602 &3.2830 \\
		\hline
		\multirow{3}{*}{1.0} & $\left \{ C,D \right \} $ & 0.5632 &  0.9088& 1.6928 &2.5138 & 3.3280  \\
		& $\left \{ C,L \right \} $ & 0.3733 & 0.6747 & 1.7251 &\underline{3.1400} & \underline{4.6024}  \\
		& $\left \{ D,L \right \} $& \underline{1.1324} & \underline{1.3852} & \underline{1.9097} &2.4296 & 2.9384 \\
		\hline
		\multirow{3}{*}{10.0} & $\left \{ C,D \right \} $ & 0.4870 & 0.6666 &1.0001 &1.3335&1.6669\\
		& $\left \{ C,L \right \} $ & 0.6676 & \underline{1.2905} &\underline{2.4584} &\underline{3.6250} &\underline{4.7917} \\
		& $\left \{ D,L \right \} $& \underline{1.0134} & 1.1424  &1.4005 &1.6585 &1.9165 \\
		\hline
	\end{tabular}
	\label{tab:2}
\end{table*}

\clearpage
\begin{table*}[htbp]
	\caption{Tournament winners at various introspection strengths with seven strategy sets. Similar to Table \ref{tab:2}, but now encompassing all strategy sets.}
	\centering
	\begin{tabular}{|c|c|c|c|c|c|c|} 
		\hline
		\multirow{2}{*}{ $w$ }& \multirow{2}{*}{space} &\multicolumn{5}{c|}{\textbf{combined score}} \\ 
		\cline{3-7} 
		& &$b=1.5$&$b=2$&$b=3$&$b=4$& $b=5$ \\
		\hline
		\multirow{7}{*}{0.1} 
		& $\left \{ C \right \} $ &-0.6922 & 0.4513 &2.7912 &5.1998  &7.6751  \\
		& $\left \{ D \right \} $ &\underline{3.9758} &\underline{5.1021} &\underline{7.3547} &\underline{9.6073} &\underline{11.8599} \\
		& $\left \{ L \right \} $&1.7500  &1.7500 &1.7500 &1.7500 &1.7500 \\
		& $\left \{ C,D \right \} $&1.7639  &2.9035  &5.2085  &7.5470  &9.9180  \\
		& $\left \{ C,L \right \} $&0.6234  &1.2181  &2.5030  &3.9172  &5.4620 \\
		& $\left \{ D,L \right \} $&2.9526  &3.5920  &4.9417  &6.3840 &7.9159 \\
		& $\left \{ C,D,L \right \}$&1.8086 &2.6090  &4.2846 &6.0573 &7.9231 \\
		\hline
		\multirow{7}{*}{1.0} 
		& $\left \{ C \right \} $ &-1.6276 &-0.6495 &1.6995 &4.2680  &6.8634   \\
		& $\left \{ D \right \} $ &3.1515 &3.9666 &5.5970 &7.2274 &8.8577 \\
		& $\left \{ L \right \} $&1.7500  &1.7500 &1.7500 &1.7500 &1.7500 \\
		& $\left \{ C,D \right \} $&1.9080  &2.8438  &4.8335  &6.8717  &8.9030   \\
		& $\left \{ C,L \right \} $&1.0809  &1.8274  &4.1137 &6.9468  &\underline{9.8203}  \\
		& $\left \{ D,L \right \} $&\underline{3.1069}  &\underline{3.9551}  &\underline{5.7315} &\underline{7.4868} &9.1944 \\
		& $\left \{ C,D,L \right \} $&2.2868 &3.1849  &5.2078 &7.3189 &9.4013 \\
		\hline
		\multirow{7}{*}{10.0} 
		& $\left \{ C \right \} $ &-2.7688 & -1.7502 &0.2503 &2.2504  &4.2505 \\
		& $\left \{ D \right \} $ &2.4621 &2.9622 &3.9622 &4.9623 &5.9623 \\
		& $\left \{ L \right \} $&1.7500  &1.7500 &1.7500 &1.7500 &1.7500 \\
		& $\left \{ C,D \right \} $&2.3669  &3.0970  &4.5209  &5.9446  &7.3683  \\
		& $\left \{ C,L \right \} $&1.7908  &3.0383  &\underline{5.4253}  &\underline{7.8105}  &\underline{10.1957} \\
		& $\left \{ D,L \right \} $&\underline{3.3280}  &\underline{3.9953}  &5.3299  &6.6645 &7.9992 \\
		& $\left \{ C,D,L \right \} $&3.0742 &3.8909  &5.5023 &7.1134 & 8.7244\\
		\hline
	\end{tabular}
	\label{tab:3}
\end{table*}

\clearpage

\clearpage

\clearpage
\clearpage

\clearpage

\end{document}